\documentclass{amsart}

\usepackage{amssymb,amsfonts,amsmath,amsthm}
\usepackage[dvips]{graphicx}
\usepackage{subfigure}
\usepackage{helvet}
\usepackage{t1enc}
\usepackage{latexsym}
\usepackage{mathrsfs}
\usepackage{color}
\usepackage{algorithm}
\usepackage{algorithmic}
\usepackage{placeins}

\graphicspath{{../}}

\usepackage[colorlinks]{hyperref}

\theoremstyle{plain}
\newtheorem{thm}{Theorem}

\newtheorem{lem}{Lemma}
\newtheorem{prop}{Proposition}

\theoremstyle{definition}
\newtheorem{defn}{Definition}

\theoremstyle{remark}

\newcommand{\etc}{\textit{etc}. }

\newcommand{\ie}{\textit{i.e.} }

\newcommand{\RR}{\mathbb{R}}
\newcommand{\NN}{\mathbb{N}}
\newcommand{\ZZ}{\mathbb{Z}}

\newcommand{\N}{\mathcal{N}}
\newcommand{\A}{\mathcal{A}}
\newcommand{\B}{\mathcal{B}}
\newcommand{\C}{\mathcal{C}}

\newcommand{\wt}[1]{\widetilde{#1}}

\newcommand{\mc}[1]{\mathcal{#1}}
\newcommand{\ms}[1]{\mathscr{#1}}

\renewcommand{\bar}{\overline}

\newcommand{\barint}{\kern4pt \raise3.4pt\hbox{\vrule height.6pt
    width7pt} \kern-11pt \int}

\newcommand{\red}[1]{\textcolor{red}{#1}}

\begin{document}

%\conflictofinterest{Conflict of interest footnote placeholder}
%\track{This paper was submitted directly to the PNAS office.}

%\footcomment{Abbreviations:}

\title{The Landscape of Complex Networks}

\author{Weinan E}
\address{Weinan E, Beijing International Center for Mathematical Research, Peking University, Beijing 100871, P.R. China; 
Department of Mathematics and Program in Applied and Computational Mathematics, Princeton University, Princeton, NJ 08544-1000 U.S.A.}
\email{weinan@math.princeton.edu}

\author{Jianfeng Lu}
\address{Jianfeng Lu, Courant Institute of Mathematical Sciences, New York University, 251 Mercer St., New York, NY, 10012-1185, U.S.A.}
\email{jianfeng@cims.nyu.edu}

\author{Yuan Yao}
\address{Yuan Yao, School of Mathematical Sciences, LMAM and LMP, Peking University, Beijing 100871, P.R. China}
\email{yuany@math.pku.edu.cn}

%\author{Weinan E\affil{1}{Beijing International
%    Center for Mathematical Research, Peking University, Beijing
%    100871, P.R. China}\affil{4}{Department of Mathematics and Program in
%    Applied and Computational Mathematics, Princeton University,
%    Princeton, NJ 08544-1000 U.S.A.}, Jianfeng Lu\affil{2}{Courant Institute of
%    Mathematical Sciences, New York University, 251 Mercer St., New
%    York, NY, 10012-1185, U.S.A.}, \and Yuan Yao\thanks{To whom
%    correspondence should be addressed. E-mail:
%    yuany@math.pku.edu.cn}\affil{3}{School of Mathematical Sciences,
%    Peking University, Beijing 100871, P.R. China}}

%\footlineauthor{E, Lu and Yao}

%\contributor{Submitted to Proceedings of the National Academy of Sciences of the United States of America}

\begin{abstract} 
  Topological landscape is introduced for networks with functions
  defined on the nodes. By extending the notion of gradient flows
  to the network setting, critical nodes of different indices are
  defined. This leads to a concise and hierarchical representation of
  the network. Persistent homology from computational topology is used to
  design efficient algorithms for performing such
  analysis. Applications to some examples in social and biological
  networks are demonstrated, which show that critical nodes carry
  important information about structures and dynamics of such
  networks.
%   The concept of local minima/maxima and saddle points of
%  different indices are defined, by extending the notion of gradient
%  flows and minimum energy path to the network setting.  Persistent
%  homology is used to design efficient numerical algorithms for
%  performing such analysis.  Applications to some examples of social
%  and biological networks are demonstrated.  These examples show that
%  the critical nodes play important roles in the structure and
%  dynamics of such networks.
\end{abstract}

\maketitle

%\begin{article}

%\begin{abstract} 
%  Topological landscape is introduced for networks with functions
%  defined on the nodes. By extending the notion of gradient flows
%  to the network setting, critical nodes of different indices are
%  defined. This leads to a concise and hierarchical representation of
%  the network. Persistent homology from computational topology is used to
%  design efficient algorithms for performing such
%  analysis. Applications to some examples in social and biological
%  networks are demonstrated, which show that critical nodes carry
%  important information about structures and dynamics of such
%  networks.
%%   The concept of local minima/maxima and saddle points of
%%  different indices are defined, by extending the notion of gradient
%%  flows and minimum energy path to the network setting.  Persistent
%%  homology is used to design efficient numerical algorithms for
%%  performing such analysis.  Applications to some examples of social
%%  and biological networks are demonstrated.  These examples show that
%%  the critical nodes play important roles in the structure and
%%  dynamics of such networks.
%\end{abstract}

%\keywords{network | landscape | critical node  | gradient flow | attraction basin | saddle | persistent homology | discrete Morse theory}

%\dropcap{N}
Networks have become ubiquitous tools for describing structures that occur in a variety of fields in the past ten or fifteen years, including biology, social sciences, economics and engineering. 
%\dropcap{O}ne of the most important revelations during the last ten or fifteen
%years is the recognition that networks are universal tools for
%describing structures that occur in many different fields, including
%biology, social studies, economics and engineering. Network structure
%is also one of the simplest structures that one can give to a set of
%data. This makes network studies one of the main components in data
%science. In fact, almost all discrete structures can be viewed as
%networks.
To study a network, one has to endow it with some mathematical
structure. The simplest mathematical structure on a network is the
graph structure. This gives rise to notions such as degrees, paths,
connectivity, etc. The distinctions between scale-free networks and
small-world networks, for example, can be studied by examining this structure,
see for example
\cite{WatStr98,Bar99,Strogatz01,ChuLu06}. But one can endow a network
with more sophisticated structures, such as geometric structure as in
the theory of manifold learning,
\cite{isomap,lle,SL00,BN03,hessian,diffmap}, or topological structure
as in the theory of persistent homology
\cite{EdeLetZom02,Ghrist07,Herbert08,Carlsson09}. These structures
allow us to probe more deeply into the nature of the network.

%In this paper, we discuss how one can endow a network with a landscape
%in a very simple and natural way. 
In this paper, we discuss how one can endow a network with a \emph{landscape}
when we study a function on the node set. The concept of landscape has been crucial in physics and chemistry in
describing complex systems, such as energy landscape \cite{Wales03}. The introduction of such a concept into
complex networks may equip us with a concise description of global
structures of networks and help explain certain dynamics such as
information diffusion and transition pathways. Many complex networks in real world carry flows of
information, products, power, etc., which are driven by local gradients of a scalar or energy \cite{Tor04,Tor08}. 
For example traffic flows may be driven by congestion function, heat flows are driven by temperature. In biomolecular
folding, conformational changes are driven by the free energy of states. 
On internet, user's attention may be driven by the centrality or significance of 
websites such as PageRank. In these cases, communities or groups emerge as metastable sets of gradient-based dynamics or energy basins. 
Therefore understanding the landscape of such functions
will be crucial to disclose associated dynamics in complex networks.
% brain networks, fMRI signals are important for disclosing the functional activities of
%neural systems. In high dimensional statistics, random variables with
%sparse (inverse) covariance structures provide plenty of examples of
%functions on graph nodes. 
%from the equilibrium distributions of Markov chain on graphs, as in
%Google's PageRank,  the logarithm of the equilibrium density defines an
% analog of free energy on the network.

In the core of the landscape lies the notion of critical nodes. In continuous setting this
meets the classical Morse theory in the study of manifolds
\cite{Milnor63}, where critical points can be located by vanishing gradients and their indices can be decided
by dimensionality of the unstable manifold passing through. However such an approach can not be applied to
the graph settings as there is no unambiguous definition of dimensionality in general. Precisely, consider an undirected graph $G=(V,E)$ with a
function defined on the node set $h:V\to \RR$.  The question we will
attempt to address  is: \emph{given a function on its nodes,
  how can we endow the network with a landscape, so that one can
  distinguish critical nodes such as the local minima, local maxima,
  and saddles}?
  
There are several studies in the literature which may lead to critical
nodes for graphs by carrying Morse theory to discrete settings. Nevertheless, none of them gives a satisfied answer
to the question. In
computational geometry one may embed the graph into a 2D-surface and then
apply Morse theory for 2-manifolds \cite{EdeHarZor03}. However, such a
surface embedding is not natural for general graphs in biological and
social networks. Another candidate is discrete Morse theory
\cite{forman98}, which studies functions defined on
all faces of cell complexes and is therefore hard to use in the graph setting above. 
A related subject is the extension of the Poincare-Hopf theorem to the graph setting,
e.g. in \cite{Knill12}. 
%which is however in pursuit of a graphical Poincare-Hopf theorem}. 
%proposed a definition of critical
%nodes of graphs based on Euler characteristics of lower neighborhood
%for the purpose of a graph version of Poincare-Hopf Theorem, which has
%however a different motivation and does not fit our settings.
 
In this paper we present a purely combinatorial approach which
starts from a discrete gradient flow induced by the function on graph
nodes.  Such an approach does not need a surface embedding, and turns
out to be closely related to persistent homology in computational
topology \cite{EdeLetZom02,Herbert08} and discrete Morse theory
\cite{forman98} without studying functions on high dimensional cells.
In particular, given a function (often referred to as an energy
function) on a network, we will define a discrete gradient flow
associated with that function, as well as minimum energy paths between
two disjoint sets of nodes.  This allows us to define critical nodes or saddles.  
Roughly speaking,
critical nodes are associated with minimum energy paths between node
pairs: index-$0$ critical nodes are simply local minima; index-$k$
critical nodes are the highest energy transition nodes of minimal
energy paths connecting index-$(k-1)$ critical nodes. 

Such a critical node analysis, as we show by examples in social networks
and biological networks, leads to a
concise representation of networks while preserving some important
structural properties. In short, the local minima
or maxima together with their attraction basins can be interpreted
as communities or groups in networks; saddle
points act as transition states between different critical points of lower
indices. In particular, in social networks index-1
saddles act as hubs in connecting communities; in biomolecular dynamics,
index-1 saddles play roles as intermediate or transition states connecting misfolded and native states. In the latter, such
an analysis does not rely on commonly used Markov state model, whence can be applied to much more general data analysis. Moreover, this approach leads to a hierarchical classification 
of nodes in the network and a global visualization of networks adaptive to the landscape of given energy function.

In algorithmic aspect, critical nodes in this paper can be computed
at a polynomial time cost with an algorithm based on computational topology by monitoring topological changes over energy evel sets, and in nondegenerate
case an almost linear algorithm exists which is scalable for the analysis of large scale networks.

% We call it \emph{topological landscape of
%   networks} in this paper.

\section{Landscape and Critical Nodes}

%To define the landscape, we start with a function on the network.  In
% many applications, there are a variety of natural and useful functions
%to consider.  For example, in brain networks, fMRI signals are
%important for disclosing the functional activities of neural
%systems. In high dimensional statistics, random functions with sparse
%(inverse) covariance structures provide plenty of examples of
%functions on graph nodes.  In addition, a large class of examples
%arise from the equilibrium distributions of random walks on graphs, as
%in Google's PageRank or the free energy distribution of protein
%folding/transition networks.

%Let us consider an undirected weighted graph $G=(V,E,W)$, where $V$ is
%the set of vertices or nodes, $E$ is the set of edges, and $W$ are the
%weights on the edges. One can associate a Markov chain with $G$ as
%follows. From each node $x \in V$, let $\N (x)$ be the set of
%neighbors of $x$, i.e. $\N(x) =\{y \in V, w_{xy} > 0 \}$.  A random
%walker will jump from $x$ to a node $y \in \N(x)$ with probability
%proportional to $w_{xy}$.  The associated transition probability
%matrix for this random walker is defined by $ P = (P_{xy}), P_{xy} =
%w_{xy}/\sum_{y\in \N(x)} w_{xy}$. If the graph is connected, there is
%a unique equilibrium distribution $\pi$ for this Markov chain.
%%In the case when the weights are symmetric $w_{xy}=w_{yx}$, the Markov chain is reversible with the detailed
%%balance condition $\pi_x P_{xy} = \pi_y P_{yx}$ hold. 
%By analogy with statistical mechanics, one can define an energy
%function (the free energy) by $h_x = -\log (\pi_x), x\in V$.

\subsection{Discrete Gradient Flow}

Throughout this paper we assume that $h$ is injective
(one-to-one). Such functions are generic in the space of real
functions on $V$. One may associate a \emph{gradient flow} of $h$ on
the graph $G$, as a map $D_{h, 0}: 2^V\to 2^V$ which maps a subset of
vertices to its immediate neighbors with lower $h$ values. More
precisely, given $x \in V$, define the neighbor set of $x$ with lower
energy $\N^-(x)=\{y\in \N(x): h(y)<h(x)\}$ and
\begin{equation}
  D_{h, 0}(\{x\}) = 
  \begin{cases}
    \N^{-}(x), 
    & \text{if } \N^-(x)\neq \emptyset; \\
    \{ x \}, & \text{otherwise.}
  \end{cases}
\end{equation}
For any $X \subseteq V$, we define
\begin{equation}
  D_{h, 0}(X) = \bigcup_{x \in X} D_{h, 0}(\{x\}).
\end{equation}
Let $D^2_{h, 0} = D_{h, 0} \circ D_{h, 0}$, \etc We say that $y$ is
\emph{reachable} from $x$, denoted by $x \succ y$ or $y \prec x$, if
$y\in D^k_{h, 0}(\{x\})$ for some $k \in \NN$, \textit{i.e.}, we can
find an energy decreasing path from $x$ to $y$. 

Note
  that our construction of the gradient flow is related to, but
  different from the gradient network \cite{Tor04,Tor08}, in which
  each node is only connected to its neighbor with the lowest energy
  (\textit{i.e.}  the neighbor in the steepest descent direction). We
also remark that the gradient flow can be viewed as a ``zero
temperature'' limit of the stochastic gradient flow introduced in
\cite{YLF} in the study of network communities.

% and splits into two new clubs after a series of conflicts, marked by
% red and blue respectively. In particular, node 1 is the coach who
% wished to raise the instruction fee, while node 34 is the president
% who did not want to increase the fee. The edges between nodes
% represent the co-activities outside normal club activities such as
% classes, which indicates the personal affinity between members. Note
% that from such activities, node 9 (in red rectangular node) should
% belong to the blue group as he has more links to the members in
% president group and he also voted in favor of the
% president. However, node 9 chooses to join the new club formed by
% the coach, node 1, because he would finish his black-belt class in a
% few months when the fission happens. The new club formed by the
% president changes the rules of class and if he chose to join it,
% then he had to re-take all the classes for years which was too much
% to him to afford.

\subsection{Local minima}

The \emph{local minima} of $h$ are those vertices whose $h$ value is
no larger than the values of its neighbors.
\begin{equation}
  \C_0 = \{ x \mid h(x) \leq h(y),\ \forall\, y \in \N(x)\}.
\end{equation}
In other words, the set of local minima are precisely the maximal
vertex set of \emph{fixed points} of the gradient flow $D_{h, 0}$.

Given a local minimum $x \in V$, its \emph{attraction basin} is defined to be:
\begin{equation}
  \mc{A}_0(x) = \{ y \mid D^\infty_{h, 0}(\{y\})=\{x\} \}. 
\end{equation}
These are the points that reach the local minimum $x$ but not any
other local minima.

\emph{Boundary or separatrix} consists of those nodes which can reach
more than one local minimum following the gradient flow
\begin{equation}
  \mc{B}_0 = \{ x \mid |D^\infty_{h, 0}(\{x\})| > 1\}.
\end{equation}

It is clear by definition that we have the non-overlapping
decomposition
\begin{equation}
  V = \mc{B}_0  \bigcup \bigcup_{x \in C_0} \mc{A}_0(x) .
\end{equation}
% 
%\begin{examp}[Zachary's Karate Club Network] 
%  Figure~\ref{fig:karate}(c) shows two local minima in oval shape,
%  nodes 1 and 34, together with their attraction basins marked in red
%  and blue, respectively. Other nodes make up of the separatrix
%  between these basins.
%\end{examp}

\subsection{Index-1 critical nodes}
Our next task is to classify the nodes in $\mc{B}_0$.  We do so
according to their role in the pathways connecting the different local
minima. In particular, \emph{index-$1$ critical nodes (saddles)} are
defined as the maxima on \emph{local minimum energy paths} connecting
different local minima.

Clearly such a definition relies on the notion of \emph{local minimal
  energy paths}, which depends on the topology of the path space.
Given two local minima, we examine all the paths connecting them. If
a path $\gamma_1$ can be deformed by the gradient flow to another path
$\gamma_2$, we say that $\gamma_1$ is \emph{deformable} to $\gamma_2$.
The \emph{local minimum energy paths} are paths which cannot be
deformed by the gradient flow.

To be more precise, given two points $a, b \in V$, we define a
\emph{path} from $a$ to $b$ as $\gamma = (w_0 \cdots w_n)$ such that
$w_0 = a$, $w_n = b$, and $w_{i+1} \in \N(w_i)$ for $i = 0, \cdots,
n-1$.  We denote the collection of paths from $a$ to $b$ as
$\ms{P}_{a, b}$.

We note the following elementary lemma, whose proof is obvious.

\smallskip
\begin{lem}\label{lem:path} Let $x \succ y $, we can then find a
  path $\gamma = (w_0\cdots w_n)$ from $x$ to $y$ such that $h(w_i) >
  h(w_{i+1})$ for $i = 0, 1, \cdots, n-1$.
\end{lem} 
\smallskip

Given two paths $\gamma_1, \gamma_2 \in \ms{P}_{a, b}$, we say
$\gamma_1$ is \emph{deformable to} $\gamma_2$, if there is a map $F:
\gamma_1 \to 2^{\gamma_2}$, such that
\begin{itemize}
  \vspace*{-.5em}
\item{(reaching)} every node in $\gamma_1$ reaches some nodes in
  $\gamma_2$, \ie for any $x\in \gamma_1$, $F(x)$ is not empty and for
  each $y \in F(x) \subset \gamma_2$, $y \prec x$;
\item{(onto)} every node in $\gamma_2$ is reachable from $\gamma_1$,
  \textit{i.e.} for any $y \in \gamma_2$, there exists $x \in
  \gamma_1$, so that $y \in F(x)$, or equivalently,
  \begin{equation*}
    \gamma_2 = \bigcup_{x \in \gamma_1} F(x).
  \end{equation*}
\end{itemize}

Let $a, b$ be two local minima. We call a path $\gamma\in
\ms{P}_{a,b}$ \emph{local minimum energy path}, if it is not
deformable to any other path in $\ms{P}_{a, b}$.
  
We define the energy of a path the maximal energy traversed by the
path, \ie $h(\gamma) = \max_{y \in \gamma} h(y)$. From the definition,
if $\gamma_2$ is deformable to $\gamma_1$, we have $h(\gamma_2) \geq
h(\gamma_1)$, so in terms of energy barrier, $\gamma_1$ is a more
preferable path than $\gamma_2$.

Given a local minimum energy path, we call the node of maximal energy
on the path an \emph{index-$1$ critical node}. The set of all
index-$1$ critical nodes is denoted by $\C_1$. We will also call local
minima \emph{index-$0$ critical nodes}, and hence the notation $\C_0$.

%\smallskip
%\begin{examp}[Zachary's Karate Club Network] 
%  Figure~\ref{fig:karate}(c) shows two index-$1$ saddles, nodes 3 and
%  32, the yellow and the green diamond nodes.  Node 3 is the lowest
%  energy node connecting the local minima nodes 1 and 34 via a minimal
%  energy path $\gamma_1=(1,3,33,34)$. Node 32 links the two local
%  minima by another local minimum energy path,
%  $\gamma_2=(1,32,34)$. Clearly these two paths $\gamma_1$ and
%  $\gamma_2$ can not be deformed to any other path by gradient flow.
%\end{examp}
%
%\medskip
%
The following fact gives a characterization of index-$1$ critical
nodes. The proof can be found in the SI. 

\smallskip

\begin{prop}[Classification of index-$1$ critical nodes]\label{thm:lower}
  All local minima in $\mc{B}_0$ are index-$1$ critical nodes. The
  other index-$1$ critical nodes will reach one of the local minima in
  $\mc{B}_0$ by the gradient flow.
\end{prop}

\smallskip

\begin{figure}[htbp]
  \centering
  \includegraphics[width=0.4\textwidth]{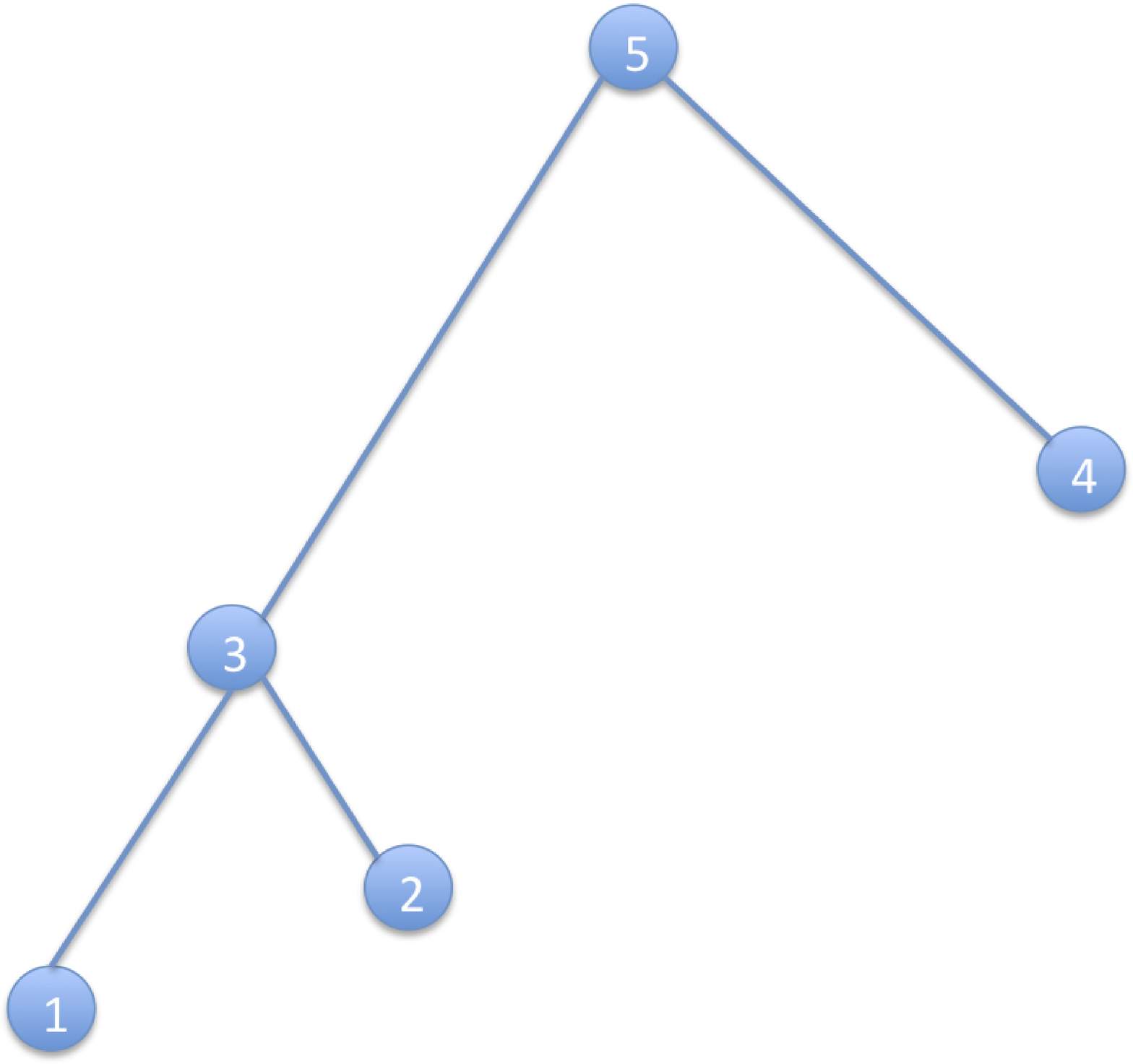}
   \includegraphics[width=0.55\textwidth]{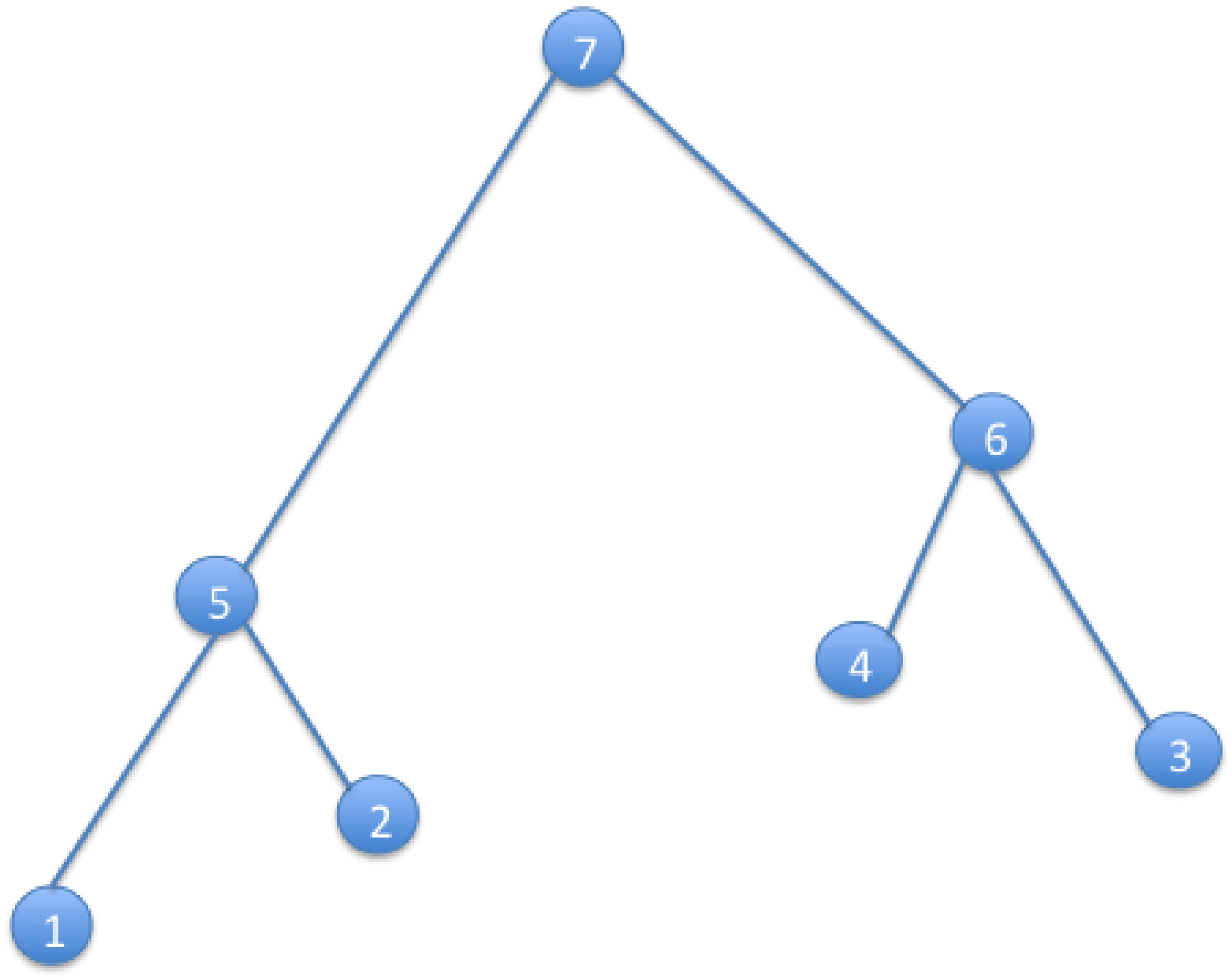}
  \caption{Left: an example of degenerate index-$1$ critical node, where node 5 on top of the tree is a degenerate index-$1$
    saddle while nodes 3 is a nongenerate index-$1$ saddle. Right: an example of both degenerate and non-degenerate critical node, where node 7 on top of the tree is a degenerate index-1 saddle as it lies on the minimum energy path connecting local minima 1 (or 2) 
and 3 (or 4), and as well a non-degenerate index-2 saddle as it is on the minimum
 energy path linking index-1 saddles 5 and 6.} \label{fig:tree}
\end{figure}

We call the index-$1$ critical nodes that are also local minima in
$\mc{B}_0$ the nondegenerate index-$1$ critical nodes, the set of
which will be denoted as $\bar{\C}_1$. The other index-$1$ critical
nodes are called degenerate. Not every index-$1$ critical node is a
local minimum in $\mc{B}_0$, for example in some cluster trees (see
Figure~\ref{fig:tree}).

\subsection{Higher index critical nodes}

The procedure presented above can be extended to define higher index
critical nodes.

To define index-$2$ critical nodes, we consider the subgraph
with nodes in $\mc{B}_0$ and edges restricted on this subset, denoted
by $G_1=(V_1=\mc{B}_0, E_1)$.  The gradient flow $D_{h, 1}:
2^{\mc{B}_0} \to 2^{\mc{B}_0}$ on $\mc{B}_0$ is defined similarly as
for $D_{h,0}$. We define the attraction basins for $x \in \mc{C}_1$ as
\begin{equation}
  \mc{A}_1(x) := \{ y \in \mc{B}_0 \mid D^\infty_{h, 1}(\{y\})=\{x\} \}.
\end{equation}
Note that for any nondegenerate index-$1$ critical node, the
attraction basin is nonempty. While for a degenerate index-$1$
critical node, the attraction basin is an empty set. This explains the
notion ``degenerate'' for the critical nodes that are not local minima
in $\mc{B}_0$.

We define the boundary set as
\begin{equation}
  \mc{B}_1 = \{ x \in \mc{B}_0 \mid |D^\infty_{h,1}(\{x\})| > 1\}.
\end{equation}
As shown in Proposition \ref{thm:lower}, all local minima on
$\mc{B}_0$ are in $\mc{C}_1$. Therefore, we have the decomposition
\begin{equation}
  \mc{B}_0 = \mc{B}_1 \bigcup \bigcup_{x \in \mc{C}_1} 
  \mc{A}_1(x)=\mc{B}_1 \bigcup \bigcup_{x \in \bar{\C}_1} \mc{A}_1(x). 
\end{equation}

Analogously, we define \emph{index-$2$ critical nodes} as the maxima
on local minimum energy paths connecting different nondegenerate
index-$1$ critical nodes. It is clear that index-$2$ critical nodes, if
exist, must be in $\mc{B}_1$.

We remark that under our definition, a degenerate index-$1$ critical
node can also be an index-$2$ critical node, as shown in Figure
\red{\ref{fig:tree}}. This ambiguity is actually quite
natural from the network point of view, as these points play multiple
roles in the structure of the network. The degenerate index-$1$
critical node can lie either in the basin of a nondegenerate critical
node or link together two different nondegenerate critical nodes.

Higher index critical nodes can be defined recursively through further
decomposition of $\mc{B}_1$.  Classification for high index critical
points can be done following similar arguments as above.  Combining
these, we obtain:

\smallskip

\begin{thm}[Node Decomposition]\label{thm:decomp}
  $V$ admits the following decomposition
  \[ V = \B_0 \bigcup \bigcup_{x\in \C_0} \A_0(x) \] where
  \[ \B_{k-1} = \B_k \bigcup \bigcup_{x\in \bar{\C}_k} \A_k(x). \]
  Here $\A_k$ is the attraction basin of local minima restricted on
  the $k-1$-th boundary set $\B_{k-1}$ and $\bar{\C}_k$ is the set of
  nondegenerate index-$k$ critical nodes.
\end{thm}

\smallskip

The theorem gives us a hierarchical representation of the network
associated to the energy landscape. It actually leads to a hypergraph representation whose hypernodes are made up of critical nodes with their attraction basins. 

%\medskip
%
%\begin{examp}[Zachary's Karate Club Network] 
%  Figure~\ref{fig:karate}(c) shows the node decomposition for Karate
%  club network with each color component for a critical node and its
%  attraction basin. In particular two basins of index-$1$ saddles are
%  in yellow (nodes 3) and green (node 32). Two index-$2$ saddles,
%  nodes 25 (in light blue diamond) and 29 (in cyan diamond), which
%  connect two index-$1$ saddles via two non-deformable minimum energy
%  paths $(3,29,32)$ and $(3,28,25,32)$. Figure~\ref{fig:karate}(d)
%  further depicts a transition path analysis (see SI) from local
%  minimum node 1 to node 34, which shows two index-1 saddles capture
%  most of transition currents.
%\end{examp}

\section{Finding Critical Nodes using Persistent Homology}

\smallskip

\subsection{Persistent homology algorithm}\ 

The landscape introduced above can be naturally formulated in terms of
a flooding procedure, from low to high values of the height function
$h: V\to\RR$.  Flooding starts from local minima, followed by the
attraction basins.  Once the relevant index-$1$ saddle is passed,
basins of local minima are merged together. This procedure then
continues on to critical points of higher indices.

More precisely, this procedure can be described in terms of persistent
homology. Persistent homology, firstly proposed by \cite{EdeLetZom02}
and developed afterwards largely in
\cite{Ghrist07,Herbert08,Carlsson09}, is an algebraic tool for
computing the Betti numbers and homology groups of a simplicial
complex when its faces are added sequentially. To work with persistent
homology, we extend the graph $G$ into a simplicial complex up to
dimension 2, and also define a filtration which consists of such
simplicial complexes, in a spirit close to \cite{EdeHarZor03} for
PL-manifolds.

An abstract simplicial complex $\Sigma_V$ is a collection of subsets
of $V$, which is closed under deletion or inclusion, i.e. if $\sigma
\in \Sigma_V$, then $\tau \in \Sigma_V$ for any $\tau\subset \sigma$.

We define \emph{the flooding complex of network $G$ associated with the function
  $h$}, $\Sigma_{G,h} \subseteq 2^V$ as follows:
\begin{itemize}
    \vspace*{-.5em}
\item $0$-simplex: the vertex set $V$;
\item $1$-simplex: the vertex pairs $\{x,y:h(x) \leq h(y)\}$ that $x
  \prec y$, \textit{i.e.}, $y \in D_{h, 0}^k(x)$ for some $k$;
\item $2$-simplex: collections of triangles $\{x,y,z:h(x)\leq h(y)\leq
  h(z)\}$, such that $x \prec y$ and $y \prec z$.
\end{itemize}
One can similarly extend the definition above to general
$k$-simplex. However for our purpose it suffices to define up to
dimension $2$ simplices.

A filtration of flooding complex $\Sigma_{G,h}$ is a nested family
$\mathcal{F}_t \subseteq \Sigma_{G,h}$ with $\mathcal{F}_{t-1}\subset
\mathcal{F}_t$ which respects the order of deletion or inclusion in
$\Sigma_{G,h}$, i.e. if $\sigma\in \mathcal{F}_t$ and $\tau \subset
\sigma$ then $\tau \in \mathcal{F}_t$.

Assume that $h:V\to \RR$ is injective or one-to-one, which is
generically the case.  By taking the maximum over vertices, one can
extend $h$ from the vertex set to simplicies, and thus to the
simplicial complex $\Sigma_{G,h}$. For a simplex $\sigma\in
\Sigma_{G,h}$ let $h(\sigma)=\max\{h(i): i\in \sigma\}$. This implies
that a face's $h$-value is always no more than that of its associated
simplex, i.e. $\sigma \subset \tau \Rightarrow h(\sigma)\leq h(\tau)$.

A filtration $(\mathcal{F}_t:t\in\NN)$ respecting the order of $h$
can be defined in the following way:
\begin{enumerate}
    \vspace*{-.5em}
\item $\mathcal{F}_0 = \emptyset$;
\item $\#\{\sigma\in \mc{F}_{t+1}\backslash
  \mc{F}_t:\dim(\sigma)=0\}=1$, \ie there is precisely one node being
  added into the filtration for each step;
\item $h(\mathcal{F}_t) < h(\mathcal{F}_{t+1})$, where $h(\mc{F}_t)
  =\max\{h(\sigma):\sigma\in \mc{F}_t\}$, \ie when a node is added
  into the filtration, all the simplices of the same energy are added
  into the filtration simultaneously.
\end{enumerate}
Note that under this construction, $\mathcal{F}_1$ consists of the
global minimum of $f$.

In this construction, we consider the filtration corresponding to the
flooding procedure from low to high $h$ values. The change of Betti
numbers identifies the index-$0$ and index-$1$ critical nodes.  Once
the filtration is defined, persistent homology computes the Betti
numbers of the simplicial complex in $\mc{F}_t$ for each $t\in \ZZ$,
and draws the barcodes of Betti number versus the $t$ or $h$ values,
{\it{e.g.}} using JPLEX
toolbox\footnote{\texttt{http://comptop.stanford.edu/programs/}}. The
proof of the following theorem is in the SI.

\medskip
\begin{thm} \label{thm:persist1}Consider the filtration
  $(\mathcal{F}_t)$. For all $t \in \NN$,
  $\mathcal{F}_{t+1}\backslash\mathcal{F}_{t}$ contains an index-$0$
  critical node if and only if $\beta_0$ increases from
  $\mathcal{F}_{t}$ to $\mathcal{F}_{t+1}$;
  $\mathcal{F}_{t+1}\backslash\mathcal{F}_{t}$ contains an index-$1$
  critical node if and only if either $\beta_0$ decreases or $\beta_1$
  increases from $\mathcal{F}_{t}$ to $\mathcal{F}_{t+1}$.
\end{thm}
\medskip

To find higher index saddles, we restrict on the subgraph $G_k=(V_k,
E_k)$ where $V_k = \mc{B}_{k-1}= V \backslash \cup_{0\leq i \leq
  k-1}\cup_{x \in \C_i} \A_i(x)$ and $E_k$ consists of edges
restricted on $V_k$. We can analogously construct the filtration
corresponds to the flooding procedure $(\mathcal{F}_{k,t}, t \in \NN)$
on the subgraph $G_k$. Similar identification holds for higher index
saddles. 

\medskip

\begin{thm} \label{thm:persistk}Consider the filtration
  $(\mathcal{F}_{k,t})$ on subgraph $G_k$ for $k\geq 2$. For all $t
  \in \NN$ such that $\mathcal{F}_{k, t+1}\backslash\mathcal{F}_{k,t}$
  contains an index-$k$ critical node if either $\beta_0$ decreases or
  $\beta_1$ increases from $\mathcal{F}_{k,t}$ to
  $\mathcal{F}_{k,t+1}$.
\end{thm}

\medskip

Clearly our characterization of high order critical nodes above only
exploits simplicial complex up to dimension $2$, whose persistent
homology computation is recently improved to be of complexity
$O(m^{2.376})$ \cite{Dimitry-SoCG11} with $m=O(n^3)$ the total number
of simplices and $n$ the number of nodes. Such a complexity does not
suffer the curse of dimensionality as the computation of high order
Betti numbers in general.

\subsection{Efficient Search of Nondegenerate
  Saddles}\label{sec:nondeg}\

As we know from Proposition \ref{thm:lower} that nondegenerate
critical nodes are actually local minimum in sub-graphs $G_k$, this
leads to an efficient algorithm for finding nondegenerate critical
nodes. In fact, all the examples shown in this paper have only
nondegenerate critical nodes and thus can be found efficiently using
this algorithm. 

Given an injective function $h$ on the vertices, we obtain the local
mimina and nondegenerate index-$k$ saddles using
Algorithm~\ref{alg:nondegenerate}.

The bottleneck in this algorithm is in finding the attraction basins
of local minima, whose complexity can be $O(nd)$ where $n$ is the
number of vertices and $d$ is the maximum degree a node has. The total
complexity is $O(Knd)$ where $K$ is the maximum index of critical
points. The algorithm is much faster than the previous algorithm for
finding all critical nodes.

\begin{algorithm}[H]
\caption{Fast search of nondegenerate critical nodes} \label{alg:nondegenerate}
{\small \begin{algorithmic}
\STATE Sort the nodes according to $h$ in increasing order;
\STATE Set $G_0=G$;
\FOR{$k=0,\ldots,n$} 
	\FOR{$x\in V_k$ in an increasing order of $h$}
	\STATE Find neighbors of $x$ with lower energy, 
	$\N^{-}_k(x)=\{y\in \N(x)\cap V_k \mid h(y)<h(x)\}$;
    \IF{$\N^{-}_k(x) = \emptyset$}
	\STATE Add $x$ to $\bar{\C}_k$ and set the color of $x$ as its node index;
	\ELSE \IF {$\N^{-}_k(x)$ contains a single color}
	 \STATE Set the color of $x$ as the single color;
	\ELSE
	\STATE Leave the color of $x$ as blank;
	\ENDIF
	\ENDIF
	\ENDFOR 
	\RETURN: \\
	\quad (1) local minima $\bar{\C}_k$ as nondegenerate critical nodes; \\
	\quad (2) attraction basins $\A_k(x_0)$ ($x_0\in\bar{\C}_k$) as color
    components; \\
	\quad (3) boundary $\B_k$ as the blank nodes; 
    \STATE Set $G_{k+1} = (\B_k, E_{k+1})$ where $E_{k+1}$ are edges 
    restricted on $\B_k$;
\ENDFOR 
\end{algorithmic}}
\end{algorithm}

\section{Examples}

\subsection{Zachary's Karate Club Network}
\begin{figure}[htbp]
  \centering
  \subfigure[]{\includegraphics[width=0.3\textwidth]{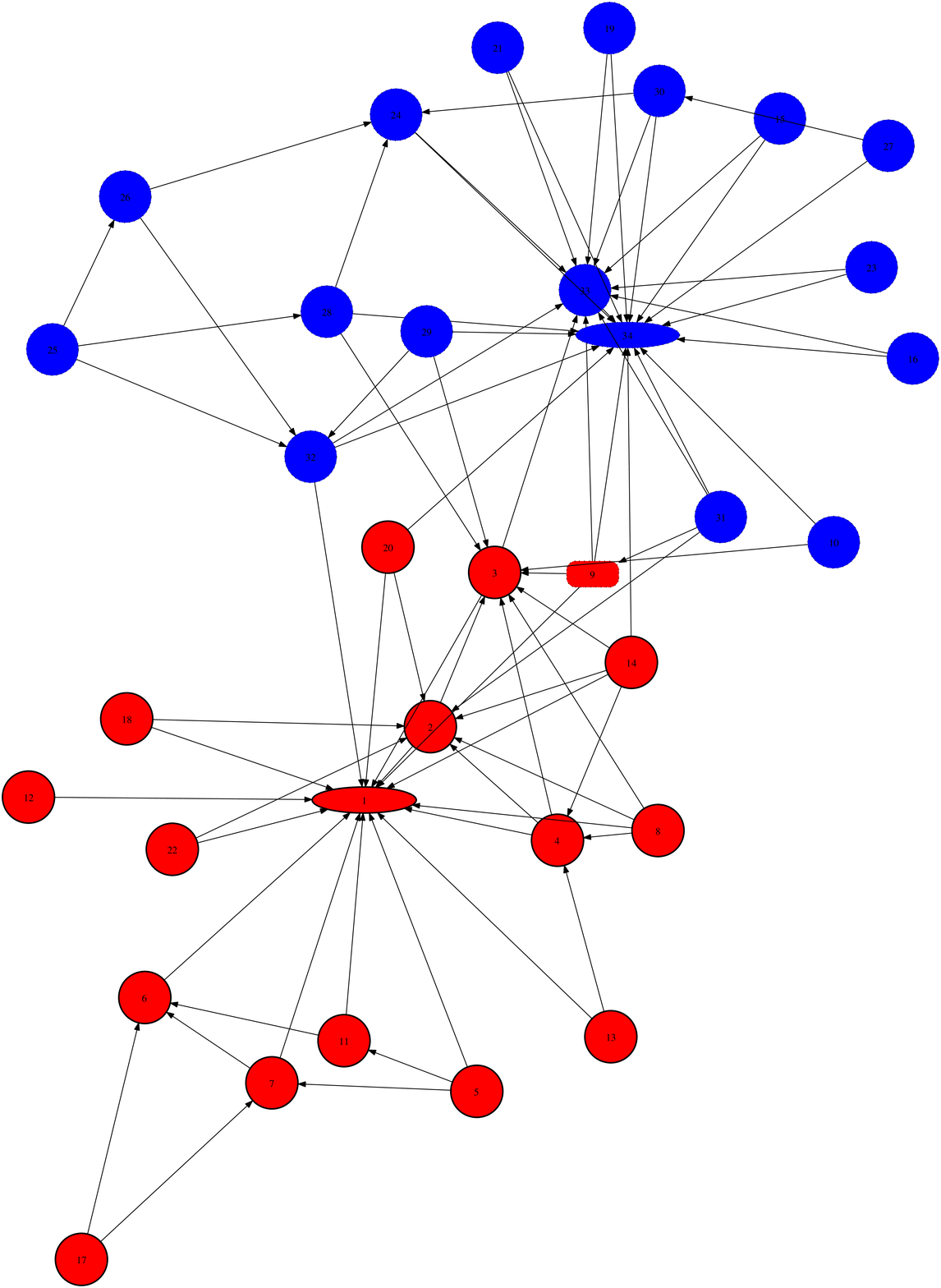}} \quad
  \subfigure[]{\includegraphics[width=0.3\textwidth]{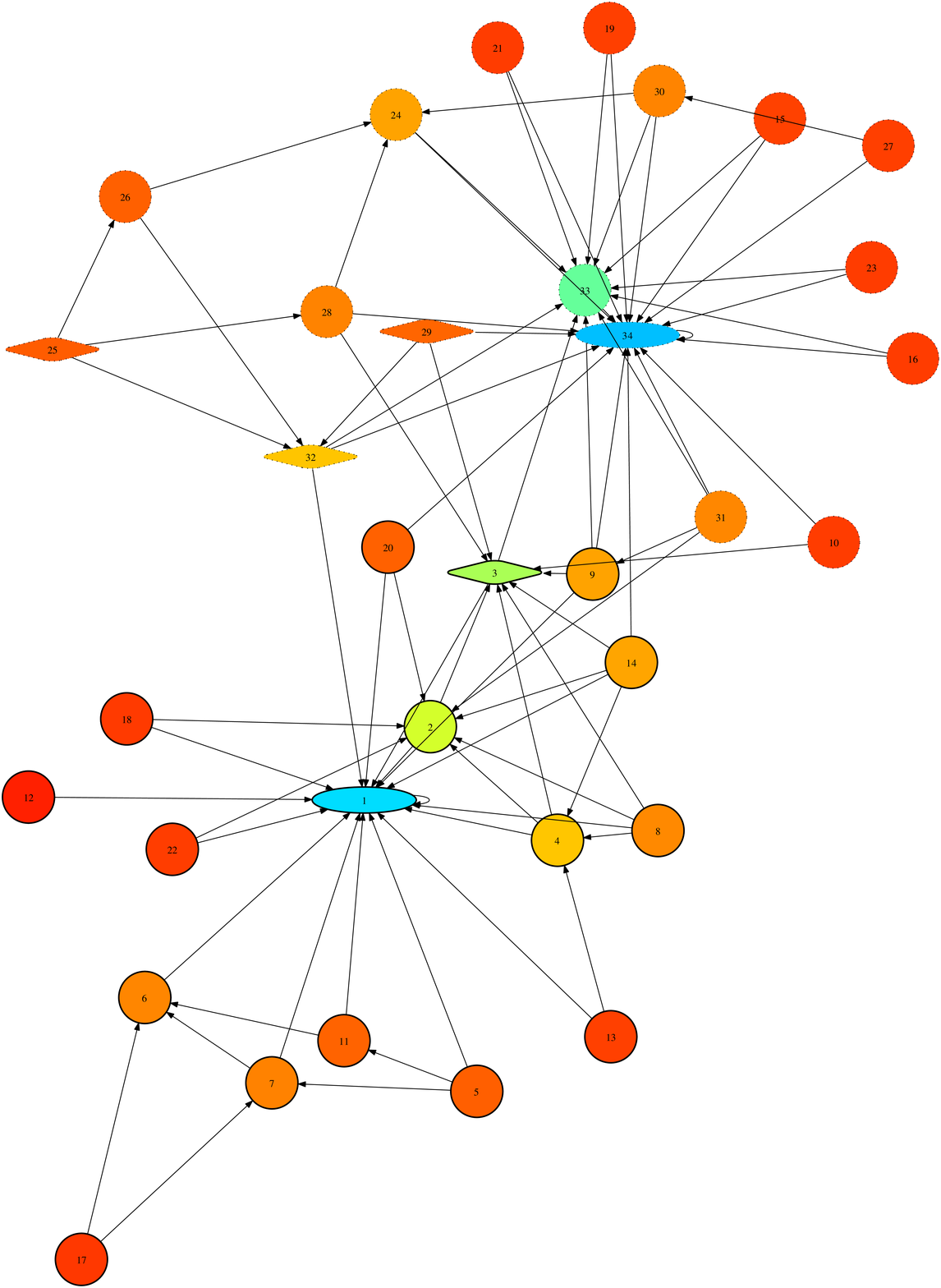}} \\
  \subfigure[]{\includegraphics[width=0.3\textwidth]{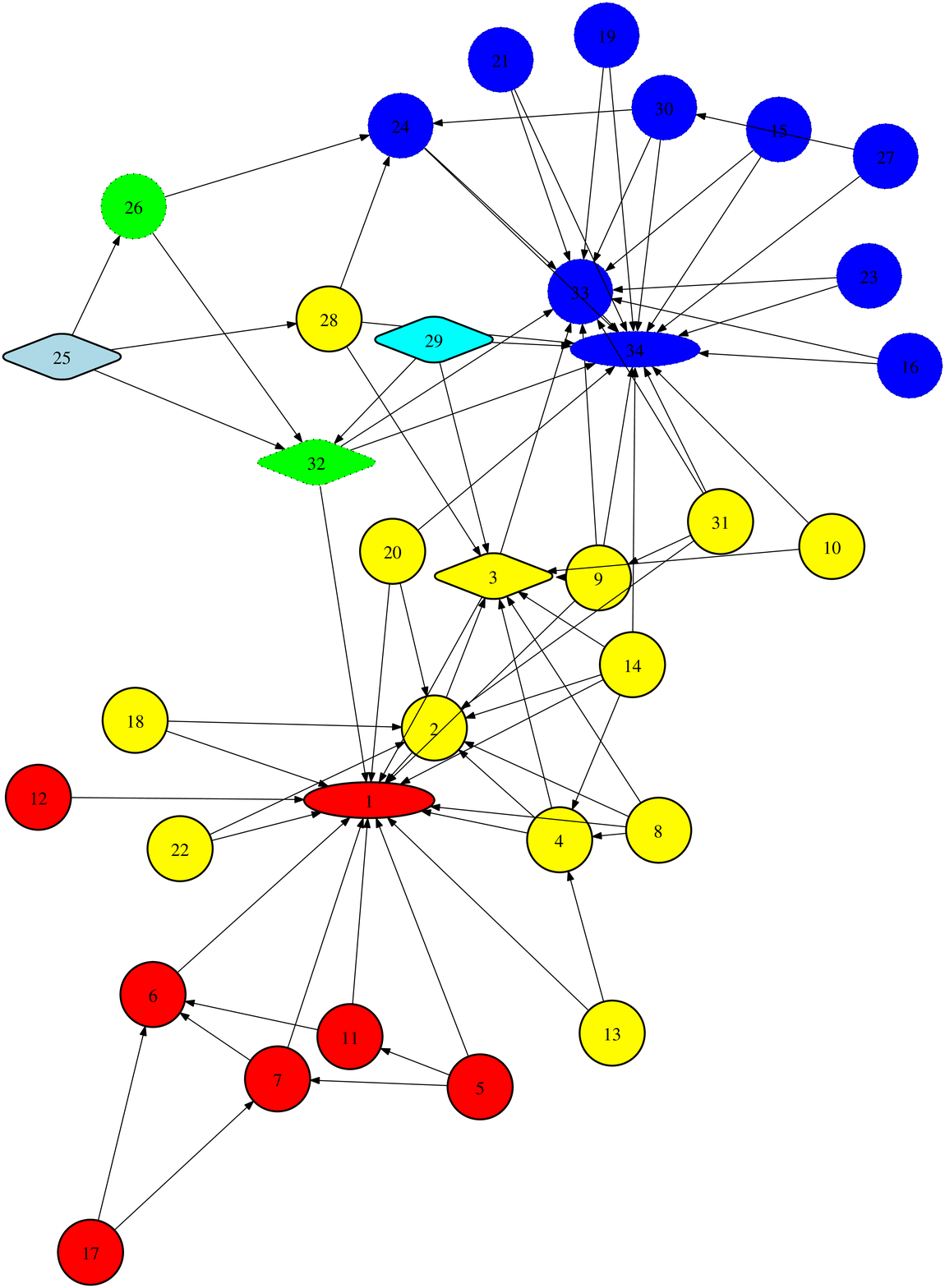}} \quad
  \subfigure[]{\includegraphics[width=0.3\textwidth]{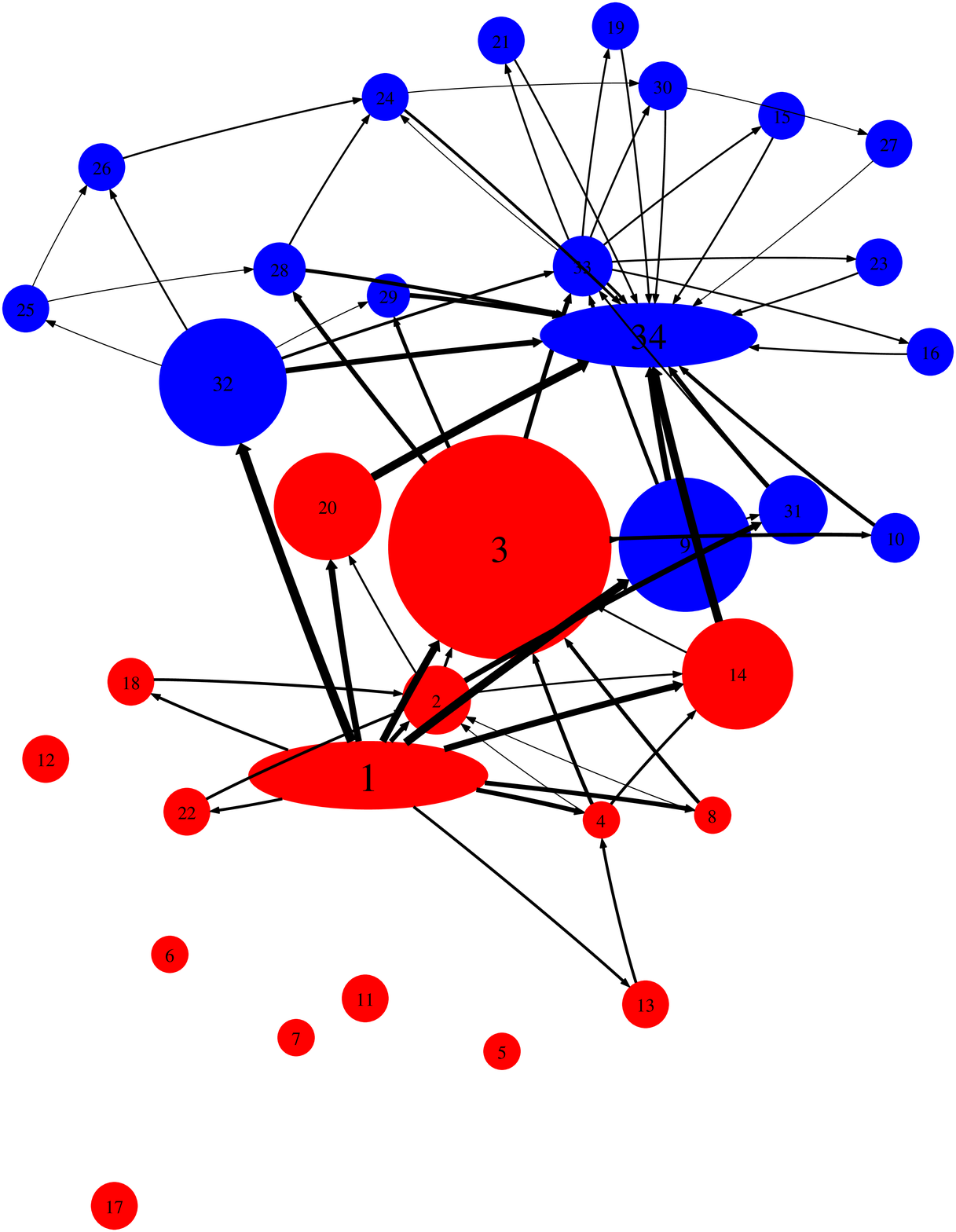}} \\
  \caption{Landscape of Karate Club. (a) The fission of Karate Club
    into two new clubs \cite{Zach77}, the coach is node 1 and the
    president is node 34, where the box node joined the red club
    (coach) instead of the blue due to his necessity to finish the
    course. (b) A gradient flow on edges, node colors from blue to red
    indicate the energy from low to high, four nodes in diamond shape
    to-be-disclosed soon as critical nodes. (c) Node decomposition
    with each color component representing a critical node with its
    basin: two local minima are in oval shape in which node 1 has
    basin in red and node 34 in blue; two index-1 saddles are in
    diamond shape in which node 3 has basin in yellow and node 32 in
    green; two index-$2$ saddles, node 25 in light blue and node 29 in
    cyan. (d) A transition path analysis (SI) with source node 1 and
    target node 34. Committor function with thresholding
    probability $0.5$ is used to divide all the nodes into two
    communities, one with node 1 in red and the other with node 34 in
    blue. Node size is in proportion to transition current connecting two communities
    through the node. Effective reactive currents from node 1 to node 34 are drawn with
    arrows on edges, whose width is determined from effective reactive
    current with a threshold greater than $0.001$. It can be seen that
    index-1 saddles (3, 32) host a majority of transition
    currents.}\label{fig:karate}
\end{figure}

Zachary's karate club network \cite{Zach77} consists of 34 nodes,
representing 34 members in a karate club with node 1 being the
instructor and node 34 being the president (Figure~\ref{fig:karate}).
An edge between two nodes means that the two members join some common
activities beyond the normal club classes and meetings. Conflicts
broke out between the instructor and the president when the instructor
sought to raise the fee and the president opposed the proposal.  The
club eventually split into two, one formed by the president (blue
nodes in Figure~\ref{fig:karate}(a)) and another one led by the
instructor (red nodes in Figure~\ref{fig:karate}(a)). A lot of
information about this fission can be disclosed by looking at the
graph structure of this social network.

\begin{figure}[htbp]
\centering
\includegraphics[width=\textwidth]{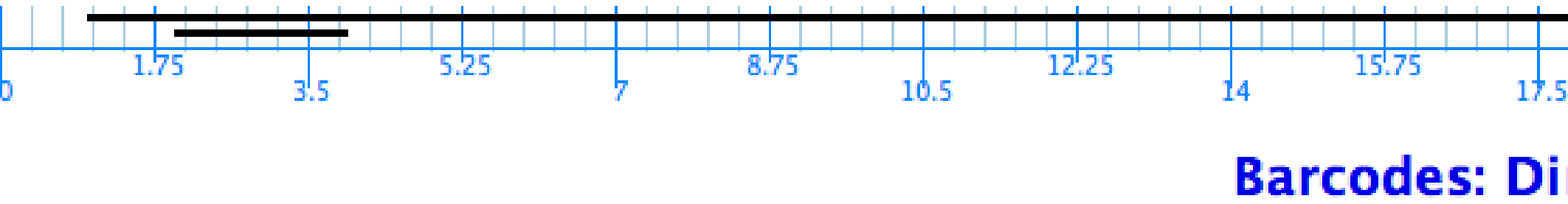} 
\includegraphics[width=\textwidth]{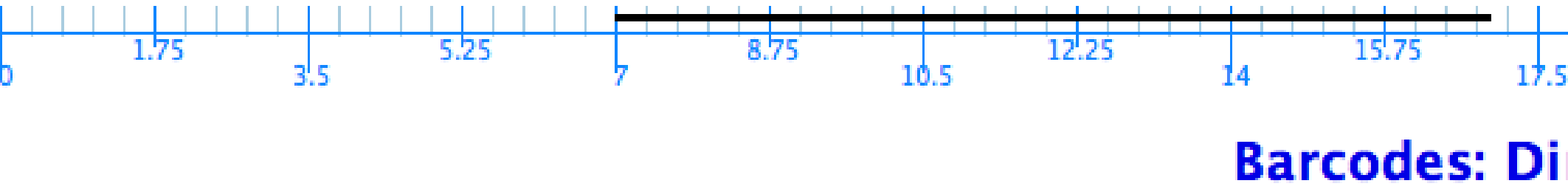} 
\caption{Barcodes of Betti numbers for the filtration of Karate Club
  network. Top: $\beta_0$ versus $t$. Node 34 with the lowest energy
  is added at $t=1$ which creates a connected component which never
  disappears. Node 1 with the second smallest energy is added at $t=2$
  which creates a new connected component disappeared when index-1 saddle 3 is
  added at $t=4$. Bottom: $\beta_1$ versus $t$. The loop is created by
  index-$1$ saddle 32 added at $t=7$ then cancelled by index-$2$ saddle 29
  at $t=17$.} \label{fig:barcode}
\end{figure}

Let $d_i$ be the degree of node $i$, and define $\bar{h}_i = -\log
d_i$. To avoid the same degree between two nodes in neighbor, a small
enough random perturbation is added such that $h_i = \bar{h}_i +
\epsilon_i$ is injective. Figure~\ref{fig:karate}(b) shows the
gradient flow of $h$. The arrows on the edges point from low degree
nodes to high degree ones. Note that nodes 24 and 25 both have degree
3, hence a small random perturbation is added resulting in the arrow
from 25 to 26. The same is done for nodes 5 and 11.

Figure~\ref{fig:karate}(c) shows the node decomposition for Karate
club network with each color component for a critical node and its
attraction basin. Two local minima, nodes 1 and 34, are in oval shape
together with their attraction basins marked in red and blue,
respectively. Two index-$1$ saddles, nodes 3 and 32, are yellow and
green diamond nodes, whose basins are in yellow (nodes 3) and green
(node 32) correspondingly. Node 3 is the lowest energy node connecting
the local minima nodes 1 and 34 via a minimum energy path
$\gamma_1=(1,3,33,34)$. Node 32 links the two local minima by another
local minimum energy path, $\gamma_2=(1,32,34)$. Two index-$2$
saddles, nodes 25 (in light blue diamond) and 29 (in cyan diamond),
which connect two index-$1$ saddles via two non-deformable minimal
energy paths $(3,29,32)$ and
$(3,28,25,32)$. Figure~\ref{fig:karate}(d) further depicts a
transition path analysis of a Markov chain induced on the graph (see
SI) from local minimum node 1 to node 34, which shows two index-1
saddles capture most of transition currents.
  
Figure~\ref{fig:barcode} shows the barcodes for the flooding complex
of this network.

\subsection{The social network of Les Mis\'{e}rables}

The social network of Les Mis\'{e}rables, collected by Knuth
\cite{Knuth93}, consists of 77 main characters in the novel by Victor
Hugo.  The edge weight $w_{ij}$ record the number of co-occurrence of
two characters $i$ and $j$ in the same scene.  Thus it is a weighted
graph where $h_i = -\log \sum_{j\sim i}w_{ij}$ as the negative
logarithmic weighted degree.  The original network exhibits a single
local (global) minimum, Valjean, who is the central character as the
whole novel was written around his experience.

 \begin{figure}[hbp]
  \centering
  \includegraphics[width=0.7\textwidth]{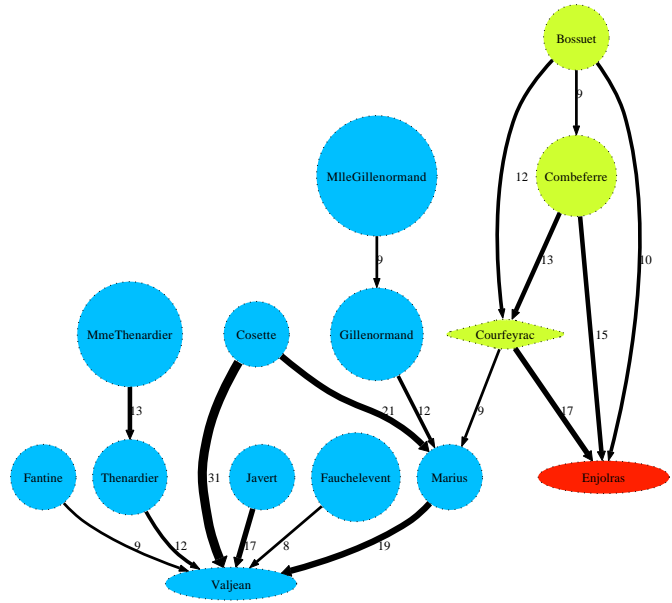} 
  \caption{Landscape of a subnet of The Les Mis\'{e}rables Network. 
    Edges are left with weights larger than $7$. Two local
    minima, Valjean and Enjoras as well as an index-$1$ saddle,
    Courfeyrac, are identified.} \label{fig:lesmis}
\end{figure}

However, dropping those edges whose weights are no more than a
threshold value ($7$ here), there appears a subnetwork which is
closely associated with the Paris uprising on the 5th and 6th of June
1832, see Figure \ref{fig:lesmis}.  The subnetwork consists of two
local minima, Enjoras and Valjean, the former being the leader of the
revolutionary students called \emph{Friends of the ABC}, the
Abaiss\'{e}.  Led by Enjolras, its other principal members are
Courfeyrac, Combeferre, and Laigle (nicknamed
Bossuet%, sometimes also written L'Aigle, Lesgle, L\`{e}gle or Lesgles
) et al., who fought and died in the insurrection.  Among them is an
index-$1$ saddle, Courfeyrac, a law student and often seen as the
heart of the group, who introduced Marius to the Friends of ABC.
Marius, a descend of the Gillenormands, though badly injured in the
battle, was saved by the main character Valjean when the barricade
fell and married to Cosette, the adopted daughter of Valjean.  The
landscape of this subnetwork highlights these events in the novel.
 
\subsection{LAO Protein Binding Transition Network}

This application examines the binding of Lysine-, Arginine-,
Ornithine-binding (LAO) protein to its ligand, recently studied in
\cite{Huang11}. The critical node analysis provides us a concise summary of global
structure of networks while preserving important pathways, which enables us to reach a more thorough description than previous approximate analysis. 

\begin{figure}[htbp]
\centering
\includegraphics[width=0.8\textwidth]{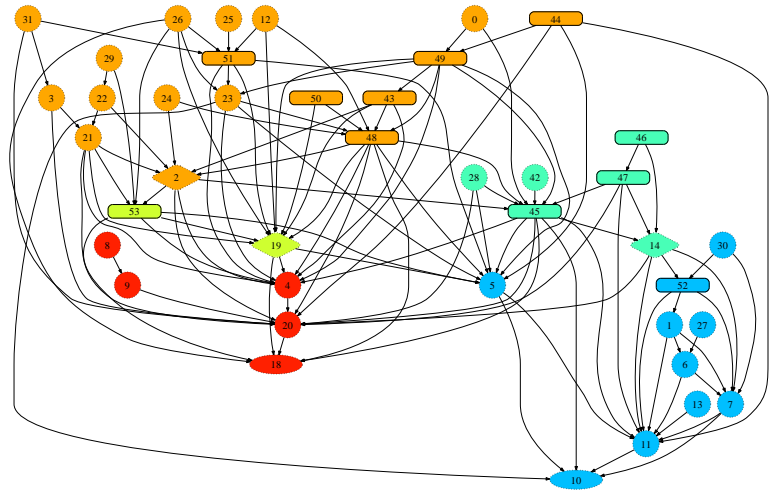} 
\caption{Landscape of LAO protein binding.  Local minima are
  represented as ovals, index-$1$ and index-$2$ saddles are shown in
  diamonds, circular nodes are regular nodes, and rectangular nodes
  are solvated states $\{43,44,\ldots,53\}$. Color components represent the node
  decomposition. }\label{fig:lao1}
\end{figure}

In \cite{Huang11} a Markov state model was constructed with 54
metastable states, using data obtained from molecular dynamics
simulation. More information about these states can be found in SI and
\cite{Huang11}. Now we examine the transition network as a weighted directed graph
$G=(V,E,W)$, where $V$ consists of 54 nodes, each representing a
metastable state, an edge $(i,j)\in E$ if transitions from node $i$ to
$j$ are observed in simulations with $6\ \mathrm{ns}$ delays (the
implied time scale for approximate Markovian behavior), and the number
of transitions is recorded as the weight $w_{ij}$.  Eleven of the
states ($\{43,44,\ldots,54\}$) are solvated or unbound states. The
binding state is node 10.
% Using transition path theory, the authors extracted the top 10
% pathways from solvated states to the binding state (node 10).

%In the ideal case that the system is reversible, $w_{ij}=w_{ji}$.   {\bf ???}
%However due to sampling errors this does not hold here. 
Let $p_{ij}=w_{ij} / \sum_j w_{ij}$ be the transition probability from
state $i$ to state $j$.  This defines a Markov chain with a unique
stationary distribution $\pi$.  We threshold this graph to an
undirected graph by keeping those edges $\{i,j\}$ such that
$\displaystyle\frac{w_{ij}+w_{ji}}{2} > 30$, \ie average count number
is larger than $30$.  One reason for doing this is that small numbers
of transitions may be heavily influenced by the noise caused by the
way of counting the transition.  Note that the mean transition count
is about $120$, and the qualitative behavior reported below shows
certain stability under the variation of the threshold value.

The energy function is $h(i)=-\log \pi(i)$ where $\pi$ is the
stationary distribution of metastable states.  Application of the
method above gives rise to a landscape shown in Figure~\ref{fig:lao1}.
Isolated states are dropped in this picture. Colors in this picture
illustrate the node decomposition according to
Theorem~\ref{thm:decomp}, where each color component represents the
attraction basin of a critical node. Below we shall discuss structural properties of 
these nodes. A complete picture of structural information for all 54 states can be
found in SI.

There are two major local minima in the landscape, nodes 10 and 18.
Node 10 is the bound state which is the minimum in the most populated
energy basin.  Its attraction basin is colored in light blue.  Nodes
11 (population $13.5\%$) and 5 (population $1.15\%$) are two encounter
complexes in that basin.  In these states, the ligand is in or close
to the binding site and conformations in this state have a small twist
but large opening angles.
% However nodes 11 and 5 play some different roles.  Once a
% conformation reaches state 11, it rapidly comes down to the bound
% state 10.  On the contrary when a conformation reaches state 5, it
% goes to 11 with a higher probability than to 10.
The other local minimum is node 18, a misbound state, where the ligand
interacts with the protein outside the binding site and close to the
hinge region of two domains of the protein. State 18, together with
state 4, 8, 9, and 20, forms a misbound basin marked in red. In these
states, the ligand interacts with the protein from a distance to the
binding site.  State 8 and 9 exhibit similar structural properties
with a negative twisting angle and a fixed distance to the binding
site (about $10{\AA}$), while state 4, 18, and 20 exhibit similar but
a different type of structures.

Node 19 and 14 are two index-$1$ saddles connecting the two basins
associated with local minima.  They are metastable intermediate states
between misbound and bound states. But in these saddles ligand
interacts with the protein in different ways. In state 19 (population
$32\%$), the ligand is interacting with the protein from one twisting
direction (positive) and the protein is quite closed. In a contrast,
in state 14 (population $35\%$) the ligand is approaching the protein
from the opposite twisting direction (negative) and the protein is
still quite open (see SI). These two saddles actually play different
roles in reactive pathways which will be discussed below. Node 2 is an
index-$2$ saddle, which is essentially a high energy misbound
state. Note that high index saddles are unstable with respect to
different thresholding values.  In the following we shall focus on
index-$1$ saddles.

For a quantitative analysis on the roles of index-$1$ saddles, we
conduct two kinds of transition path analysis using transition path
theory (\cite{E-Eric} or see SI). First, we study reactive currents from the misbound
state 18 to the bound state 10.  This analysis shows that a majority
of flux passes through the saddle 19. Therefore once the ligand and
protein fall in the misbound state 18, the major pathway to escape and
enter the bound state is via saddle 19.

The other analysis, as was also did in \cite{Huang11}, studies
transition paths from the eleven solvated states marked from 43 to 53
to the bound state 10.  In particular, we investigate reactive
currents from each of the solvated states to the bounded state,
respectively. The results are summarized as follows. A large part of
these details has been ignored in \cite{Huang11}, since they only
examined $10$ transition pathways, ignoring the others.
\begin{enumerate}
%    \vspace*{-.5em}
  \item Solvated state 52 lies in the basin of bound state 10, whence misbound state 18 has little influence on its pathway.
  \item Solvated state 53 only passes through index-$1$ critical node
    19 to enter the bound state 10, which is heavily influenced by the
    misbound state 18.
 \item Solvated states $\{45, 46, 47\}$ lie in the basin of index-1 critical node 14 and enter the bound state 10 directly or via 14. They are not much
influenced by the misbound state 18.
 \item Other solvated states are in the basin of index-2 critical node 2. Transition path analysis further shows that misbound state 18 has a stronger influence on them than those in the basin of 14. In particular state 50 is mostly influenced with near $50\%$ of transition currents trapped by the misbound state 18.  
%    than through 14. The process is  influenced by the misbound state 18. \\
%    (2) States 44, 49 pass through state 14 more than through 19. This
%    process is not influenced by the misbound state 18.
%    directly or via index-$1$ critical node 14.
%    \item Pathways from all other solvated states might or might not be trapped by misbound 18: 
%     \\ (1) States $\{45, 48\}$ are heavily influenced by misbound 18 due to direct paths to it.
%     \\ (2) States $\{44, 46, 47\}$ are not much influenced by 18 due to connections to saddle 14 or encounter complex 11.
%     \\ (3) Transition path analysis identifies that $\{49,50,51\}$ are slightly influenced by 18 while state 43 is not due to a mysterious pathway . 
%  \item Solvated state 45, 46, 47, and 52 enter the bound state 10
%    directly or via index-$1$ critical node 14. They are not much
%    influenced by the misbound state 18.
%  \item Other solvated states, 43, 48, 50, 51, 49, and 44 might be
%    influenced by both saddles: \\
%    (1) States 43, 48, 50, 51 pass through state 19 more often
%    than through 14. The process is  influenced by the misbound state 18. \\
%    (2) States 44, 49 pass through state 14 more than through 19. This
%    process is not influenced by the misbound state 18.
\end{enumerate}

In summary, the misbound state 18 affects some of the pathways from solvated states to the bound state. Index-1 critical node 14 is a state where ligand starts to interact with protein to enter the encounter complex 11. If we can design some mutations to disrupt the stability of this state or even encounter complexes, we may be able to make the binding much more difficult. Finally we note that the critical node analysis here does not rely on the Markov model assumption and can thus be applied to the analysis of transition networks in molecular dynamics beyond its Markovian time scale.  

\section{Discussion and Conclusion}

We have introduced a notion of critical points for network which can
be used to reduce a complex network to a coarse-grained
representation while preserving structural properties associated with
functional gradient flows. Examples have shown that the information
obtained this way is of great value in capturing global structure and
dynamics of the network, such as diffusive or reactive pathways.
Moreover, the critical point
analysis leads to a hierarchical decomposition which may enable us to perform
multiscale analysis of complex networks. These perspectives will be
systematically pursued in the future.

An interesting question is the stability of these objects against noise. To
answer this question, one has to clarify the source of noise. There
are two types of noise one should consider in landscape analysis of
networks -- one associated with the energy function $h$ and the
other associated with the network structure. The former can be
dealt with traditional persistent homology denoising, where critical
nodes with shallow basins can be merged with their saddles. The latter
is however more challenging as there are no systematic studies yet on
perturbation or bootstrapping of networks. In the examples above, we
used edge thresholding on the Les Mis\'{e}rables and the protein
binding networks, which is equivalent to modeling such networks as a
superposition of a signal graph and some Erd\"{o}s-R{\'{e}}nyi type
random graphs as noise. However there might be better models
which lead to different denoising rules.

%We have introduced a way of performing critical point analysis on a
%network. Examples have shown that the information obtained this way is
%of great value in visualizing the local and global structure and
%dynamics of the network, such as diffusive or reactive pathways. Given this information, one can then
%performed more refined analysis of the dynamics, for example, by examining the transitions between
%different basins with the transition path theory. The analysis is not
%limited to Markov state models, and can in principle be applied to
%more general situations, such as energy functions associated with distributions 
%on discrete structures.
%
%We have introduced a way of performing critical point analysis on a
%network. Examples have shown that the information obtained this way is
%of great value in visualizing the local and global structure and
%dynamics of the network.  Given this information, one can then
%performed more refined analysis of the dynamics, for example, by
%examining the more detailed dynamics of the transitions between
%different basins using the transition path theory. The analysis is not
%limited to Markov state models, and can in principle be applied to
%more general situations.
%

%\begin{acknowledgments}
\section*{Acknowledgements}
  W.E. acknowledges supports from ARO grant W911NF-07-1-0637 and ONR
  grant N00014-01-1-0674.  J.L. is grateful to Eric Vanden-Eijnden for
  helpful discussions. Y.Y. thanks Xuhui Huang for providing Figure
  S-2 in supporting information with helpful discussions, as well as
  supports from the National Basic Research Program of China (973
  Program 2011CB809105), NSFC (61071157), Microsoft Research Asia, and
  a professorship in the Hundred Talents Program at Peking University.
%\end{acknowledgments}

\bigskip

\renewcommand{\thesection}{S-\arabic{section}}
\renewcommand{\theequation}{S-\arabic{equation}}
\renewcommand{\thefigure}{S-\arabic{figure}}

\setcounter{section}{0}

\section*{Supplementary Information}
\section{Proofs}

\begin{proof}[Proof of Proposition 1]

  We show first that every local minimum in $\B_0$ must be an index-$1$
  critical node. Let $x$ be a local minimum in $\mc{B}_0$. Then $x$
  reaches at least two local minima, say $y_1,y_2 \in\C_0$. Consider
  the subgraph with node set
  \begin{equation*}
    S = \bigl(\{x\}\cup \A(y_1) \cup \A(y_2)\bigr) \cap \{ y
    \mid h(y) \leq h(x) \}.
  \end{equation*}
  Clearly, $S$ is connected and $x$ is the unique maximum node in $S$.
  By the definition of the attraction basin, the set $S \backslash \{x
  \}$ is not connected.

  Since $S$ is connected, it contains at least a path from $y_1$ to
  $y_2$. Let $\gamma$ be the local minimal energy path from $y_1$ to
  $y_2$ in the subgraph $S$. As $S \backslash \{x\}$ is not connected,
  $\gamma$ must pass $x$, so that $h(\gamma) = h(x)$.

  We now show by contradiction that $\gamma$ is also a local minimal
  energy path in the original graph $V$. Suppose we can find another
  path from $y_1$ to $y_2$, called $\wt{\gamma}$, so that $\gamma$ is
  deformable to $\wt{\gamma}$. For any $z \in \wt{\gamma}$, we have
  $h(z) \leq h(x)$. Consider the set $\wt{\gamma} \cap \mc{B}_0$,
  which is non-empty. We distinguish two cases:
  \begin{enumerate}
    \vspace*{-.5em}
  \item[a)] $\wt{\gamma} \cap \mc{B}_0 = \{ x \}$. Then, $\wt{\gamma}
    \backslash \{x \} \subset \mc{A}(y_1) \cup \mc{A}(y_2)$, so that
    $\wt{\gamma} \subset S$. By construction of $\gamma$, we have
    $\wt{\gamma}=\gamma$; %$\textcolor{blue}{h(\wt{\gamma}) = h(\gamma)=h(x)}$.
  \item[b)] If there exists $z \in \wt{\gamma} \cap \mc{B}_0$ and $ z \neq
    x$, we have some point $x' \in \gamma$ that $z \prec x'$. It is
    easy to see that $x'$ must be $x$, since other points on $\gamma$
    are in attraction basins of $y_1$ and $y_2$. %Using Lemma~\ref{lem:path}, 
    Using Lemma 1, there exists a path $\gamma_1 = (w_0 \cdots
    w_n)$ from $z$ to $x$ ordered in energy increase. In particular, consider
    the point $w_{n-1}$, we have $z \prec w_{n-1}$ so that $w_{n-1}
    \in \mc{B}_0$. Moreover, $w_{n-1} \in \N(x)$ and $h(w_{n-1}) <
    h(x)$. This contradicts with the fact that $x$ is a local
    minimizer in $\mc{B}_0$.
  \end{enumerate}
  Therefore, $\gamma$ is a local minimal energy path, and $x$ is an
  index-$1$ critical node. 

  Let $z \in \mc{C}_1$ which is not a local minimum in
  $\mc{B}_0$. Then, $z$ must reach a local minimum $x$ in $\mc{B}_0$
  by the gradient flow. By the first part of the proposition, $x \in
  \mc{C}_1$. The proposition is proved. 
\end{proof}

\begin{proof}[Proof of Theorem 2]
  
  (Necessity). We first show that index-$0$ and index-$1$ critical
  nodes, when added into the filtration, will change Betti numbers in
  the way above.
  
  For index-$0$ critical nodes, they are local minima of graph
  $G$. When a local minima is added into the filtration, it must
  create a new connected component which increases the $0$-th Betti
  number, $\beta_0$.

  Index-$1$ saddles will play a more complicated role. We have two
  situations
  \begin{itemize}
    \vspace*{-.5em}
  \item if an index-$1$ saddle lies on top of a global minimal energy
    path, it will decrease $\beta_0$ upon being added;
  \item if an index-$1$ saddle lies on top of a local minimal energy
    path other than the global one, it will increase $\beta_1$ upon
    being added.
  \end{itemize}
 
  Given a pair of index-$0$ critical nodes $y_1, y_2\in \C_0$, among
  all local minimal energy paths connecting them (if exist), there
  must be a global minimal energy path $\gamma_0$, so that
  $h(\gamma_0)$ is less than any other local minimal energy paths
  between $y_1$ and $y_2$. We denote the maximal node of the global
  minimal energy path as $x$. Such $x$ is an index-$1$ critical
  node. When $x$ is added into the filtration, the $0$-th Betti number
  $\beta_0$ will decrease as $x$ connects two components contains
  $y_1$ and $y_2$ respectively. 

  For the other local minimal energy paths connecting $y_1$ and $y_2$,
  the associated index-$1$ critical nodes will increase the first
  Betti number $\beta_1$ when added into the filtration. Indeed, let
  $z$ be such an index-$1$ critical node. Thus $z$ is a maximum of a
  local minimum energy path $\gamma_1$ such that
  $h(\gamma_1)=h(z)>h(x)=h(\gamma_0)$. $\gamma_1$ is not deformable to
  the global minimal energy path $\gamma_0$ between $y_1$ and
  $y_2$. Then two paths $\gamma_0$ and $\gamma_1$ forms a loop, and
  hence the first Betti number $\beta_1$ increases when $z$ is added
  into the filtration.

  \medskip

  (Sufficiency). We show next that no other nodes when added into the
  filtration will change the first two Betti numbers in the same way.
  
  For any node $x$ which lies in the attraction basin of a local
  minima $\A_0(x_0)$ for some $x_0\neq x$, $x$ reaches $x_0$ by
  gradient flow. For any edge $\{x,x^\prime\}\in E$ with $x^\prime \in
  \A_0(x_0)$, $x^\prime$ reaches $x_0$ and thus the triangle
  $\{x,x^\prime,x_0\}$ is included in the simplicial complex. This
  implies that $\A_0(x_0)$ is contractible (star-shape), whence no
  node in $\A_0(x_0)$ other than local minimum $x_0$ will change Betti
  numbers.

  It remains to show that any node in boundary $\B_0 \backslash \C_1$
  will not change Betti numbers in the same way. Any such node $z\in
  \B_0$ must reach at least two local minima, say $a$ and $b$. Then by
  Lemma \ref{lem:path} there is a path $\gamma=(a= w_0, \ldots, z=
  w_k,\ldots, b= w_l)$ for some $l\in \NN$ such that $h(w_s)<
  h(w_{s+1})$ for $s\leq k-1$ and $h(w_s)> h(w_{s+1})$ for $s>k$.
  Moreover $z\not\in\C_1$ implies that $\gamma$ is deformable to a
  local minimal energy path $\pi = (a = v_0, \ldots, b = v_m)$ between
  the same end nodes, for some $m\in \NN$. $z$ can not decreases
  number of connected components as the path $\pi$, which appears
  first in the filtration, already connects $a$ and $b$.

  Now we show that the path $\gamma$ will not create a loop
  either. Let $\pi_t=c \in \C_1$ be the maximal node on $\pi$. We must
  have $c \prec z$. To see this, as $\gamma$ is deformable to $\pi$,
  there is a node $c^\prime = w_{k^\prime}\in \gamma$ which reaches
  $c\in \pi$. We may assume $c^\prime \neq z$ ($k^\prime \neq k$)
  since otherwise we are done. Then, by the construction of the path
  $\gamma$, we have $c' \prec z$, and hence $c \prec z$. 
  
  Note that both $z$ and $c$ reach both local minima $a$ and $b$, node
  $w_i$ with $i<k$ ($i>k$) reaches $a$ ($b$, respectively), and node
  $v_i$ with $i<t$ ($i>t$) reaches $a$ ($b$, respectively). These will
  create a set of triangles such that $\gamma$ is homotopy equivalent
  to $\pi$, \ie loop-free.
\end{proof}

\begin{proof}[Proof of Theorem 3]
 The proof is analogous to that of Theorem 2. %Theorem~\ref{thm:persist1}.
\end{proof}

\section{Current on edges and paths. Transition path theory}

The energy landscape gives us a global picture for the different
attraction basins on the network.  To understand the dynamics between
the different basins, the transition path theory (TPT) provides a
natural tool.

The transition path theory was originally introduced in the context of
continuous-time Markov process on continuous state space \cite{E-Eric}
and discrete state space \cite{MeScVa09}, see \cite{EVa10} for a
review. Another description of discrete transition path theory for molecular dynamics
can be also found in \cite{Noe09}. Here we adapt the theory to the setting of discrete time
Markov chain with transition probability matrix $P$. We assume
reversibility in the following presentation, the extension to
non-reversible Markov chain is straightforward.

Given two sets $A$ and $B$ in the state space $V$, the transition path
theory tells how these transitions between the two sets happen
(mechanism, rates, etc.). If we view $A$ as a reactant state and $B$
as a product state, then one transition from $A$ to $B$ is a reaction
event. The reactve trajectories are those part of the equilibrium
trajectory that the system is going from $A$ to $B$. To make the
notion more precise, define the ordered family of times $\{n_j^A,
n_j^B\}$ such that
\begin{align*}
  &X_{n_j^A} \in A, \quad, X_{n_j^B} \in B,  \\
  &X_{n} \in V\backslash(A \cup B), \quad \forall n, n_j^A < n <
  n_j^B.
\end{align*}
Hence, a reaction happens from time $n_j^A$ to time $n_j^B$.

\medskip
\begin{defn} Given any equilibrium trajectory $\{X_n\}$, we call each
  portion of the trajectory of between $n_j^A$ and $n_j^B$ a
  \emph{$AB$-reactive trajectory}. We call the time during which the
  reaction occurs the \emph{reactive times}
  \begin{equation}
    R = \bigcup_{j\in \ZZ} (n_j^A, n_j^B).
  \end{equation}
\end{defn}
\medskip

The central object in transition path theory is the committor
function. Its value at $x$ gives the probability that a trajectory
starting from $x$ will hit the set $B$ first than $A$, \textit{i.e.},
the success rate of the transition at $x$.  Given two sets $A$ and $B$
in the state space, $q$ satisfies the equation
\begin{equation}
\begin{cases}
  \sum_{y \in V} p_{xy}q(y) - q(x)  = 0, & x\not\in A\cup B; \\
  q(x) = 0, & x\in A; \\
  q(x) = 1, & x\in B,
\end{cases}
\end{equation}

The committor function provides natural decomposition of the graph. If
$q(x)$ is less than $0.5$, $x$ is more likely to reach $A$ first than
$B$; so that $\{ x \mid q(x) < 0.5\}$ gives the set of points that are
more attached to set $A$. 

Once the committor function is given, the statistical properties of
the reaction trajectories between $A$ and $B$ can be quantified. We
state several propositions characterizing transition mechanism from
$A$ to $B$. The proof of them is an easy adaptation of \cite{E-Eric,
  MeScVa09} and will be omitted.

\medskip
\begin{prop}[Probability distribution of reactive trajectories] 
  The probability distribution of reactive trajectories
  \begin{equation}
    \pi_R(x) = \mathbb{P}(X_n = x, n \in R) 
    % = \lim_{T \to \infty} \frac{1}{2T} \sum_{n \in R}\delta(X_n = i),
  \end{equation}
  is given by
  \begin{equation}
    \pi_R(x) = \pi(x) q(x) ( 1 - q(x)).
  \end{equation}
\end{prop}
\medskip

The distribution $\pi_R$ gives the equilibrium probability that a
reactive trajectory visits $x$. It provides information about the
proportion of time the reactive trajectories spend in state $x$ along
the way from $A$ to $B$. 

\medskip
\begin{prop}[Reactive current from $A$ to $B$]
  The reactive current from $A$ to $B$, defined by
  \begin{equation}
    J(xy) = \mathbb{P}(X_n = x, X_{n+1} = y, \{n, n+1\} \subset R),
  \end{equation}
  is given by 
  \begin{equation}
    J(xy) = 
    \begin{cases}
      \pi(x) q(x) P_{xy}(1 - q(y)), & x\neq y;\\
      0, & \text{otherwise}.
    \end{cases}
  \end{equation}
\end{prop}
\medskip

The reactive current $J(xy)$ gives the average rate the reactive
trajectories jump from state $x$ to $y$.  From the reactive current,
we may define the effective reactive current on an edge and transition current
through a node which characterizes the importance of an edge and a
node in the transition from $A$ to $B$, respectively.

\medskip
\begin{defn}
  The \emph{effective current} of an edge $xy$ is defined as
  \begin{equation}
    J^+(xy)=\max(J(xy)-J(yx),0).
  \end{equation}
  The \emph{transition current} through a node $x \in V$ is defined as
  \begin{equation}
    T(x) = \left\{\begin{array}{ll}
    \sum_{y \in V} J^+(xy) ,& x\in A \\
    \sum_{x\in V} J^+(xy), & x\in B \\
    \sum_{y\in V} J^+(xy) = \sum_{x\in V} J^+(xy), & x\not\in A\cup B  
    \end{array}\right.
  \end{equation}
\end{defn}
\medskip

In applications one often examines partial transition current through a node connecting two communities $V^-=\{x:q(x)<0.5\}$ and $V^+=\{x:q(x)\geq 0.5\}$, \emph{e.g.} $\sum_{y\in V^+} J^+(xy)$ for $x\in V^-$, which shows relative importance of the node in bridging communities. 

The reaction rate $\nu$, defined as the number of transitions from $A$
to $B$ happened in a unit time interval, can be obtained from adding
up the probability current flowing out of the reactant state. This is
stated by the next proposition.

\medskip
\begin{prop}[Reaction rate]
  The reaction rate is given by
  \begin{equation}
    \nu = \sum_{x\in A, y\in V} J(xy)
    = \sum_{x\in V, y\in B} J(xy).
  \end{equation}
\end{prop}
\medskip

Finally, the committor functions also give information about the time
proportion that an equilibrium trajectory comes from $A$ (the
trajectory hits $A$ last rather than $B$).

\medskip
\begin{prop}
  The proportion of time that the trajectory comes from $A$
  (resp.~from $B$) is given by
  \begin{equation}
    \rho^{A} = \sum_{x\in V} \pi(x) q(x), \quad 
    \rho^{B} = \sum_{x\in V} \pi(x) (1 - q(x)).
  \end{equation}
\end{prop}
\medskip

%\FloatBarrier

\section{Supplementary Figures} 

\setcounter{figure}{0}

\begin{figure}[htbp]
\centering
\includegraphics[width=0.9\textwidth]{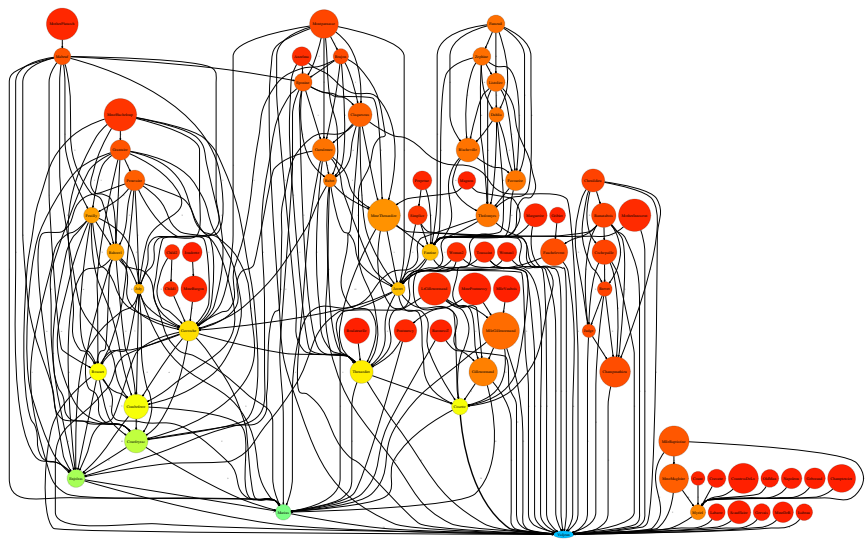} 
\caption{The Les Mis\'{e}rables Network. The whole
  network has 77 nodes as main characters in Victor Hugo's novel, Les
  Mis\'{e}rables, where Valjean is the only local minimum (global
  minimum) as the novel is written around his experience. } 
  \label{fig:lesmis}
\end{figure} 

\begin{figure}[htbp]
\centering
\includegraphics[width=0.9\textwidth]{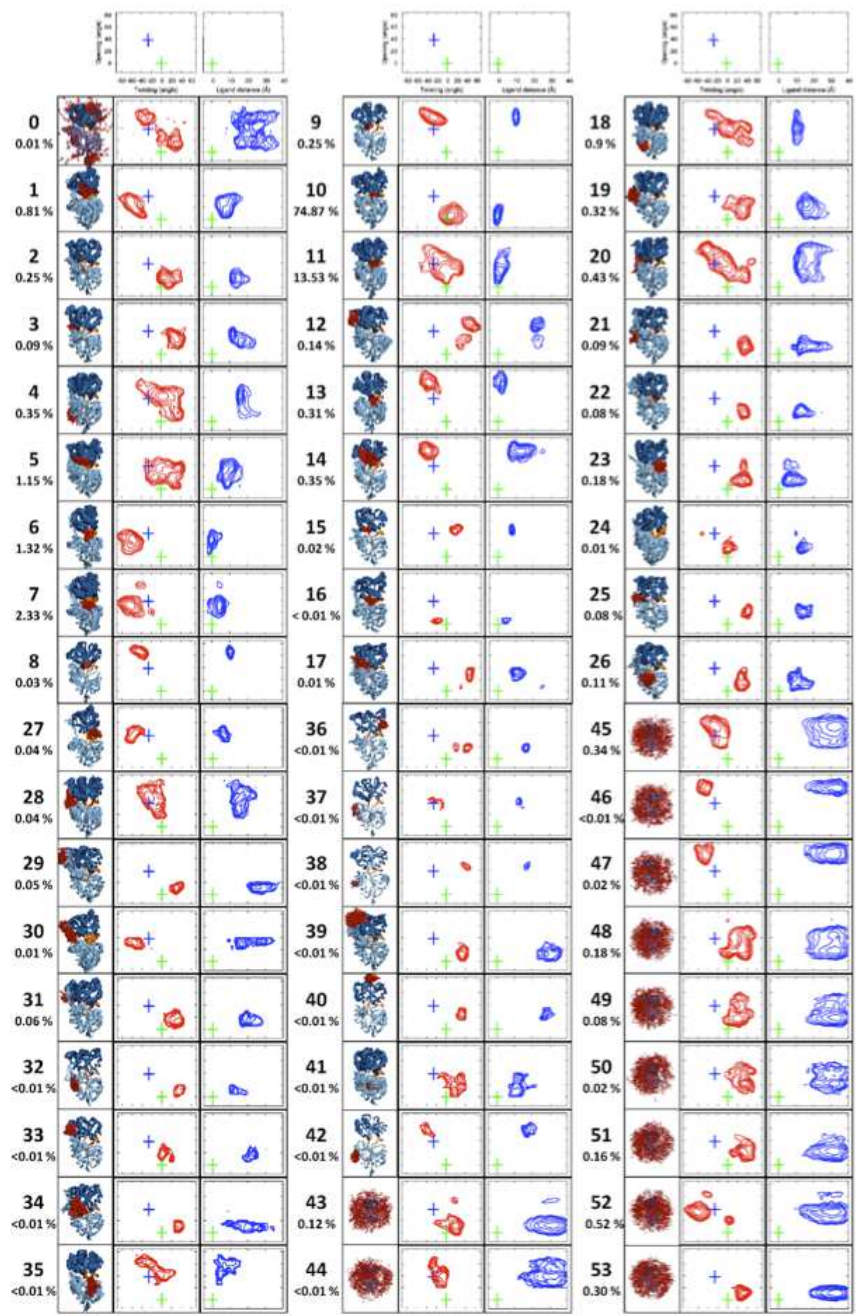} 
\caption{(Courtesy by Xuhui Huang) Three types of pictures for each of the 54 metastable states:
  on the left is the crystal structure of a representative
  conformation in each state, on the right are free energy plots of
  the protein opening angle versus twisting angle (O, T) (red), as
  well as the distance between the ligand and the binding site versus
  the opening angle (L,O) (blue).  The green and blue crosses
  correspond to X-ray structures of the bound (PDB ID: 1LAF) and apo
  (PDB ID: 2LAO) conformations respectively. }\label{fig:lao2}
\end{figure} 

\begin{figure}[htbp]
\centering
\includegraphics[width=0.8\textwidth]{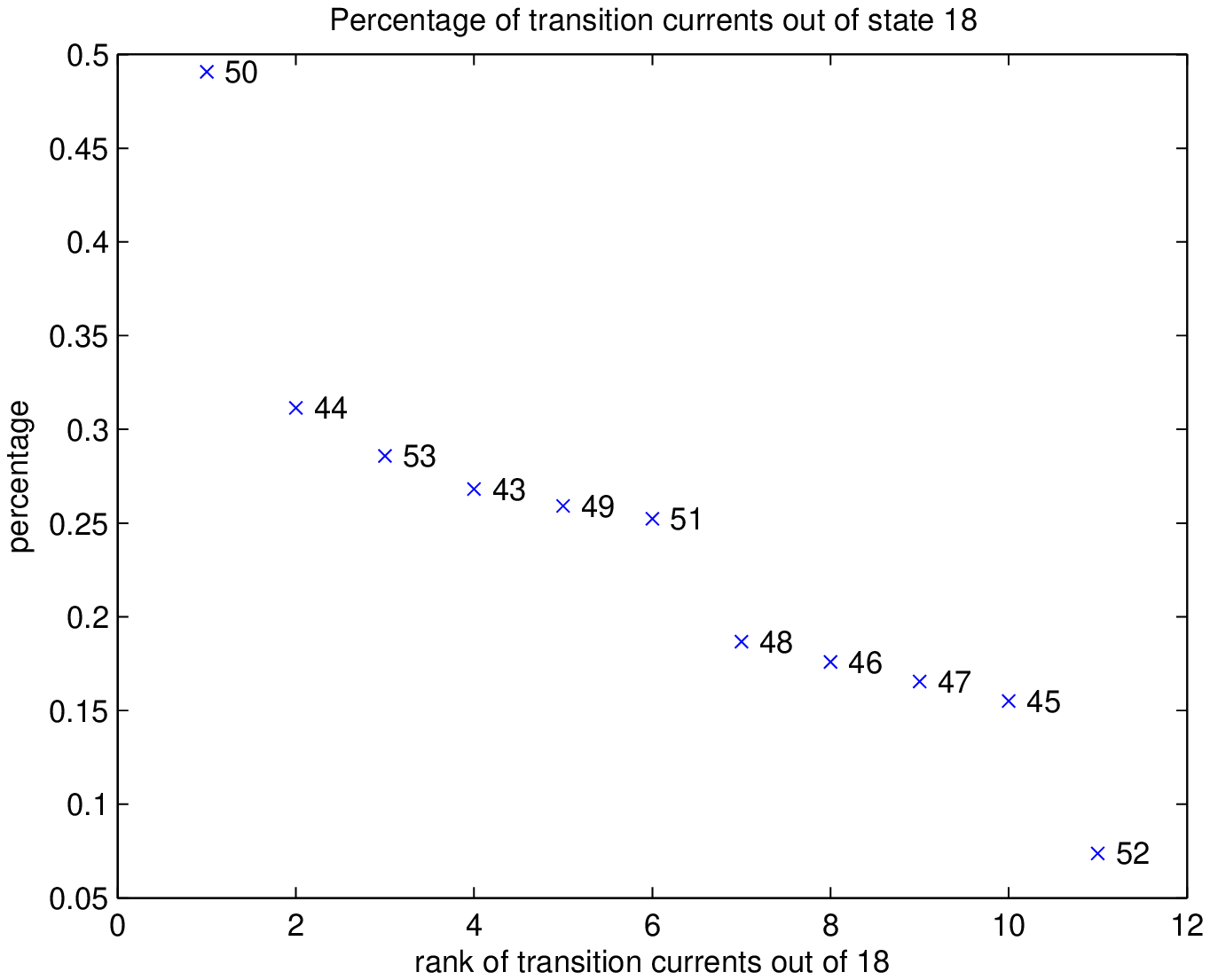} 
\caption{Transition Currents out of misbound state 18, with source set from each of solvated states $\{43,\ldots,53\}$
and target set as bound state $10$.}\label{fig:lao-trancurrent}
\end{figure} 

The first figure is the whole co-appearance network of 77 main
characters in the novel, Les Mis\`{e}rables, by Victor Hugo
\cite{Knuth93}. It is an undirected weighted graph with edge weights as
the number of co-appearances for a pair of characters. Without thresholding
this network contains one local minimum, Valjean. However a thresholding with 
edge weight greater than 7 gives rise to the subnetwork in the main text.

The second figure contains a list of structural information on 54
metastable states. It contains a typical crystal structure in each state, and
some free energy plots on certain reaction coordinates. From these pictures 
one can read various structural properties of critical nodes in LAO-protein binding transition network
discussed in the main text. More information about this system can be found in \cite{Huang11}.

The third figure shows the ranking of transition currents out of misbound state 18 over eleven transition pathways.
The experiment selects each of the eleven solvated states $\{43,\ldots,53\}$ as the source set and the misbound state 10 as the common target set. In each of the eleven experiments, relative transition current out of state 18 divided by total transition current from the source, is recorded and plotted in a descending order.

\bibliographystyle{pnas}

\end{document}